# A District-level Flood Severity Index for Flood Management in India

Manabendra Saharia[1], Sharad K Jain[2], Ved Prakash[1], Harshul Malik[1], O P Sreejith[3], Dheeraj Joshi[4]

[1]Department of Civil Engineering, Indian Institute of Technology Delhi, Hauz Khas, New Delhi 110016, India

[2]Department of Civil Engineering, Indian Institute of Technology Roorkee, Roorkee, Uttarakhand 247667, India

[3]India Meteorological Department, Pune, India

[4]Indian Railway Service, National Mission for Clean Ganga, Department of Water Resources, RD & GR, Ministry of Jal Shakti, New Delhi, India

**Corresponding Author**:

Manabendra Saharia, Ph.D.

Associate Professor, Department of Civil Engineering

Associate Faculty, Yardi School of Artificial Intelligence

Indian Institute of Technology Delhi

New Delhi, India 110016

HydroSense Lab: https://hydrosense.iitd.ac.in/

**Abstract**

India is one of the worst affected countries in the world in terms of fatalities and economic damage due to natural disasters, particularly floods. For planning flood mitigating and relief measures, granular historical information on a pan-India basis is required, which has been missing. Through recent efforts, a few national scale datasets have been created, but they lack the requisite information on fatalities and damages, which has limited the ability to develop a flood severity index. This paper describes the development of the India Flood Inventory with Impacts (IFI-Impacts) database, which contains death and damage statistics, and combines population and historically flooded area information sourced from a national hydrologic-hydrodynamic modeling system. We also propose a novel District Flood Severity Index (DFSI), which accounts for the historical severity of floods in India based on the number of people they have affected and the spread and duration of such floods. Districts being the administrative units of the government, this novel index fulfills a major need and gap in currently available flood management tools. The dataset as well as the index is expected to significantly advance disaster preparedness towards floods in the country. DFSI can be improved further by collecting and incorporating additional variables, e.g., economic losses and by improving the reliability/robustness of the data of other variables. Based on DFSI, actions need to be addressed to mitigate flood damages, beginning with the districts with the high DFSI values.

**Highlights** (3-5 bullet points that capture the novel results of your research as well as new methods that were used. Be clear and concise and use simple terms)

1. India faces floods recurrently, causing huge adverse impacts to life and assets. However, a comprehensive flood severity index is missing.
2. A district level flood severity index is developed using the India Flood Inventory with Impacts (IFI-Impacts) database containing data of variables representing the occurrence of floods and damage due to floods.
3. Since district is the most relevant unit for planning and decision making, a DFSI would be of immense value for flood management.

## 1. Introduction

India is subjected to monsoonic climate in which about 75% of the annual rainfall takes place over the rainy season of four months (June-Sep). A large concentration of rainfall leads to recurrent floods in many parts of the country. The impacts of floods in India have been severe, with 113,390 human casualties reported between 1975 and 2015, averaging 2765 deaths per year (Saharia et al. 2021a). In addition, floods also damage infrastructure including bridges and roads, crops, building, etc., inflicting economic losses. Many efforts are being made in India to manage floods so as to protect human lives and property. To design better flood management strategies and focus attention to the most critical areas, an essential requirement is to know the places that are most vulnerable to flood damage. A flood severity map captures the spatial and temporal information in terms of extent, duration, and damage caused by floods. While flood hazard studies over India using hydrologic or statistical models are abundant in literature, no flood severity index map that incorporates ground level information at fine scales is currently available.

Preparation of flood severity map for a large and diverse country like India is challenging due to several reasons. Data related to floods belongs to a number of sectors: meteorology, hydrology, agriculture, transportation, urban, rural, revenue, health, disaster management, etc. Floods data are collected by a large number of agencies with different objectives, mandates, skills, and available tools. The formats of data collection are also widely varying. Currently, there is no national/regional scale system to compile the data which may be buried in paper records, computer files, maps, etc. Often, there is lack of data sharing and after the event or intended use/application of the data is over, it finds very little usage in long-term flood management studies.

Studies show that the spatial organization of rainfall influences flood severity along with geomorphology and climatology (Saharia et al. 2021b). India being a large country, wide variation in these causative factors is seen. In the catchments in north India (Ganga, Brahmaputra, and Indus), floods may occur due to intense rains as well as snow/glacier melt. In recent times, frequency of occurrence of Glacial Lake Outburst Floods (GLOFs) is rising; specifically, three GLOF events (Sikkim 2023, Rishiganga 2022, and Kedarnath 2013) have caused large number of deaths, and damage to infrastructure, roads, bridges, and buildings. Moreover, frequency of high intensity rainfall events is also rising, leading to severe flooding at many places. Besides the Himalayan river basins, floods also occur in the catchments of Godavari, Narmada, Mahanadi, Tapi rivers, and in the state of Kerala. Apart from these, frequent urban floods (Mumbai, Chennai, etc.) have also caused many damages and have impacted the economy.

To assess myriad impacts of floods, we require datasets that not only capture the information about the floods, but also on the consequent losses and damages. While there are national scale flood modelling systems such as the Indian Land Data Assimilation System (Magotra et al. 2024), there is a growing need for flood severity indices that are based on observational datasets. Various flood severity indices have been developed for the United States (Schroeder et al. 2016; Saharia et al. 2017; Kumar et al. 2024), Europe (Marchi et al. 2010), Greece (Diakakis et al. 2020). But in India, such studies have been limited to case studies (Ghosh and Dey 2021; Kanth et al. 2022), and a national flood severity index is missing. The risk of floods at a given place depends on the climate of the region, properties of precipitation upstream of the place, topography, soils and geology, land use/cover, and human factors/interventions.

Concerned about the flood occurrences in India, Government of India had constituted the National Commission on Floods (Rashtriya Barh Ayog, RBA) in 1976 which submitted a detailed report in 1980 (Rashtriya Barh Ayog 1980). Till date, this report is one of the most comprehensive attempts to study the problem of floods in India and continues to be used for various studies. RBA had put forth a number of recommendations and in the past, many efforts have been made to implement these. Study of floods figures prominently in the activities of the Central Water Commission (CWC), a technical organization under the Ministry of Jal Shakti, and of the Ganga Flood Control Commission, Govt. of India. CWC also operates a Flood Management & Border Areas Programme (FMBAP). Research institutes such as the National Institute of Hydrology, Indian Institute of Tropical Meteorology, National Center for Medium Range Weather Forecasting, numerous academic institutions and organizations of State Governments are also contributing to solve flood problems. In addition, different State Governments also have the ministries that deal with flood management. National Remote Sensing Centre (NRSC), an institute of the Indian Space Research Organization (IRSO), publishes flood inundation maps for different years prepared using satellite data. A number of strategies have been suggested for flood mitigation in India. A view is that flood waters may be considered as a resource to be conserved, to the extent feasible, and be used to meet the water demands in the dry summer season which comes a few months after flood (Jain and Singh 2020).

CWC had published a map in 2023 on flood prone areas of India which is based on slightly old data. Our analysis shows that the flood prone area in India has changed considerably in the recent times and this map needs to be updated. In the last few decades, many floods have occurred at places that are not considered flood prone. Recently, National Remote Sensing Centre (NRSC 2023) has released a "Flood Affected Area Atlas of India - Satellite based Study", covering period of 25 years (1998-2022) by using both optical and microwave satellite data. This atlas is closer to ground reality but also has inherent limitations as the estimated flood extent depends on the availability of satellite data, date of overpass of the satellites and available coverage over flooded areas. The NRSC atlas has presented data and analysis state wise. It describes major flood events, district-wise flood affected area, and submerged roads and railway. Optical and microwave satellite produced images are also included. But the NRSC atlas doesn't cover flood events that had lasted for short time and could not be mapped due to non-availability of satellite data for short duration floods. Thus, it is expected that there may be underestimations in flood affected area in the NRSC atlas. Further, the atlas does not provide information about lives lost and economic damages, etc. Fig. 1 shows the CWC (2013) produced map of area liable to flood and flood affected area map produced by NRSC. It is noted that the two maps have some agreement and some differences. In CWC map, most area liable to floods are located in the Indus, Ganga, and Brahmaputra basins, on the east coast, and in Gujarat. Very small or no areas in major river basins such as Narmada, Godavari, Krishna, and Cauvery are categorized as liable to floods. In the NRSC map, most flood affected area falls in Ganga and Brahmaputra basins, and in West Bengal, Odisha, Gujarat, Andhra Pradesh, and Tamil Nadu states. The area liable to flood (CWC map) is more than the flood affected area (NRSC map).

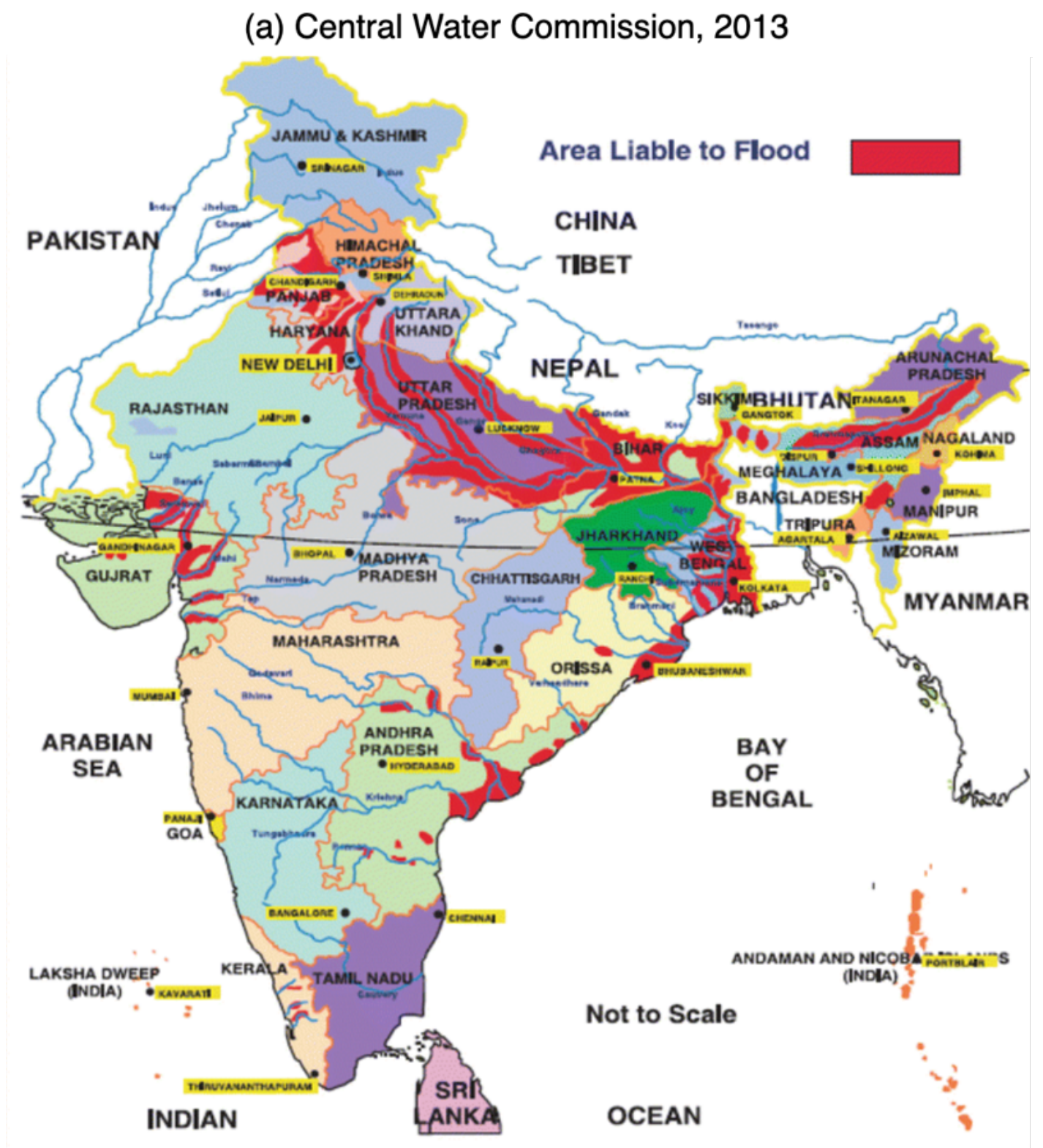

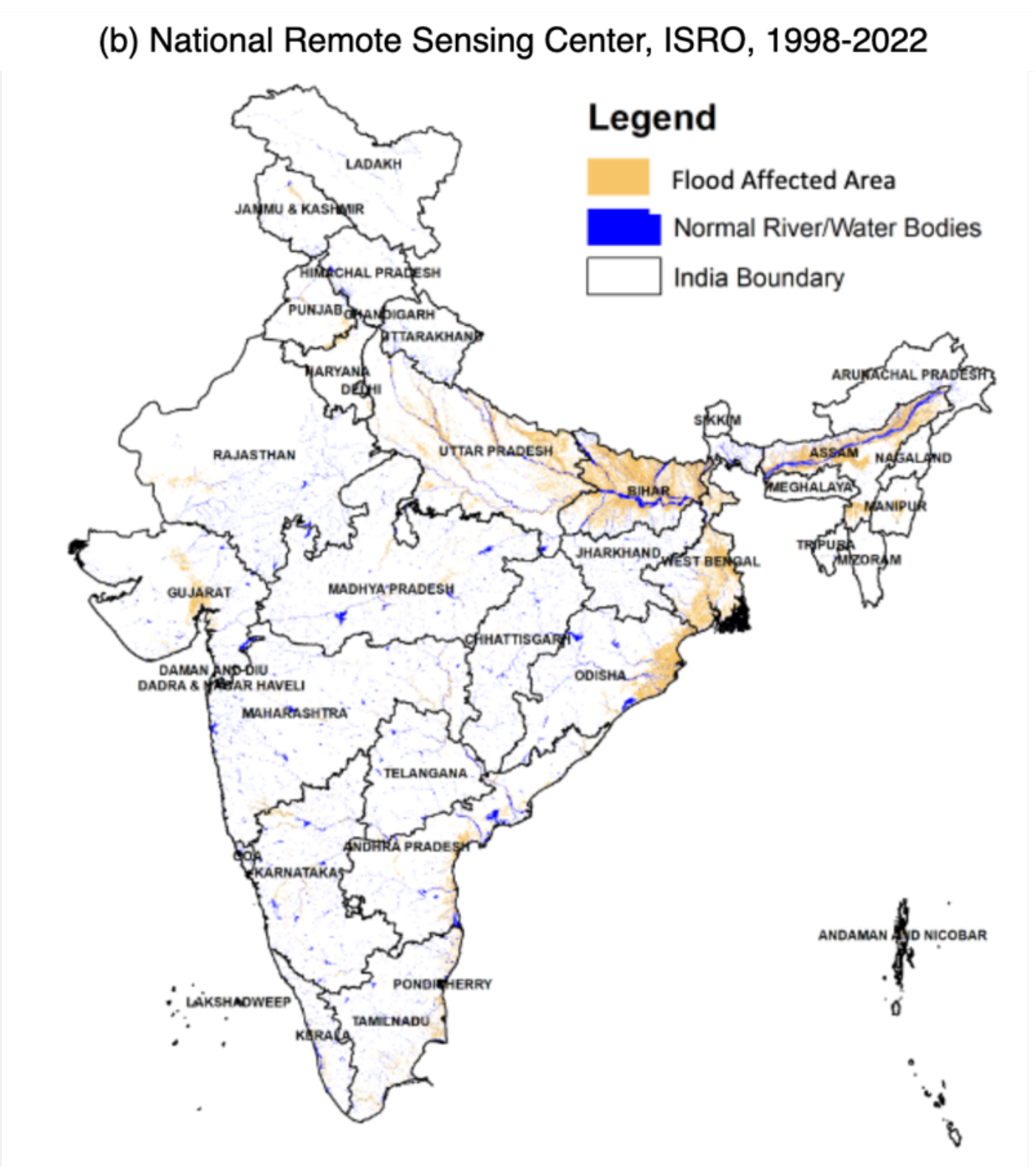

*Figure 1: (a) Map of flood prone area in India as identified by the Central Water Commission (CWC, 2013), and (b) Map of flood affected area using satellite data from 1998-2022 by the National Remote Sensing Center, ISRO.*

Despite the fact that flooding is a highly damaging disaster in India, there is a lack of comprehensive data pertaining to floods or a dedicated flood portal in India. One such effort is the building of the India Flood Inventory (IFI), which is the first freely available, analysis-ready geospatial dataset over the region with detailed qualitative and quantitative information regarding floods, including spatial extents (Saharia et al. 2021a). To draw plans for flood management, data on many other features, related to flood damages are required at finer spatial resolution. These data include, human causality due to floods, deaths of cattle, economic damages, e.g., damage to crops, roads, bridges, buildings, infrastructure, etc. Some of these data are collected by various government agencies but remain buried in official records at different locations in different departments. No inventory of these data is maintained, thereby limiting their accessibility and use. Researchers and academicians may not be aware of the existence of such data and find it hard to access and use them. Further, well-articulated protocols to collect, validate, and report these data have not been developed/followed.

This study presents an updated India Flood Inventory with Impacts (IFI-Impacts) dataset, which contains detailed death, injury, damage and flood information from 1967-2023. We also ran a national hydrologic-hydrodynamic modeling system to find historically flooded areas of the country. Finally, we have developed a novel district flood severity index (DFSI) which is based on district wise information on flood duration, fatality, injury, flooded area, and population.

## 2. Causes of Flooding in India

In India, floods are generated due to many different causes: pluvial, fluvial, snowmelt/GLOF, cyclones, and drainage congestion. Pluvial or rain-induced floods are the most common among them. Though large number of hydrogeomorphic factors influence the cause and evolution, floods are often generated by short-to-medium duration intense rainfall or moderate-to-long rainfall over large areas. In both cases, once an area receives high volume of rainfall which exceeds the capacity to infiltrate or drain out the water, it leads to inundation of the low lying areas. This becomes more complicated in mountainous areas where loose and exposed soils may be easily eroded leading to sediment-laden flows or mudflows. Drainage congestion occurs when an area receives more inflow than what can be quickly let out, resulting in piling of water and inundation. This process is often cited as a major reason for floods in many parts of India such as the Gangetic plains of in UP and Bihar states and Brahmaputra plains in Assam.

India is a developing country that is experiencing rapid population growth and large-scale changes in land use and land cover (LULC) which is altering the hydrologic responses of river basins. The LULC changes of interest include deforestation, unplanned urbanization, construction in low lying areas and wetlands, etc. When the LULC changes lead to an increased generation of overland flows, the area may experience high and frequent flooding compared to its natural condition. Now, in addition, climate change is likely to significantly alter the patterns, intensity, and duration of rainfalls in a catchment and subsequently its hydrologic response. Due to climate change, localized intense rainfall events are expected to rise in future, which could lead to more flash floods. A warming climate and oceans can also increase the frequency and intensity of cyclones which will impact the coastal and nearby inland areas with intense rainfalls and floods. In the future, new areas may become prone to flooding and this will require additional efforts to provide protection to these places.

## 3. Methodology and Data Used

The following datasets and methodology are used in this study.

### *3.1* India Flood Inventory with Impacts (IFI-Impacts)

The original IFI dataset contained information on flooding from 1967-2016, sourced from an annual printed publication named "Disastrous Weather Events" (DWE) by the India Meteorological Department (IMD), but has undergone extensive cleaning and adaptation to make it amenable for computational research. But the dataset had no district information, and the focus was flood events, rather than impacts in terms of loss and damage to life and property. The new IFI-Impacts database digitizes voluminous amount of information stuck in printed documents published by various government departments in India, which have seldom been used in research. This data is collected by IMD offices around the country and are ground-validated, which increases the trustworthiness of the data in terms of ascertaining damages, fatalities, as well as spatial extents. Apart from loss and damage data, the IFI-Impacts dataset also contains the historically flooded area of a district in percentage derived from the streamflow reanalysis provided by the Indian Land Data Assimilation System (ILDAS, (Magotra et al. 2024)). Districts are the basic administrative units in India. But, some of these have been reorganized and/or split over the time. In some instances, district names have been changed and this posed a challenge in reconciling the data. Another difficulty was that in many cases, more than one spelling of district name is used. Therefore, preparation of this database required careful screening of data and manual corrections, which was a laborious task.

### *3.2* Flooded area from the Indian Land Data Assimilation System (ILDAS)

The Indian Land Data Assimilation System (ILDAS) was developed as a hydrologic-hydrodynamic modeling system to estimate land surface states, channel discharge, and floodplain inundation over South Asia (Magotra et al. 2024; Chakraborty et al. 2024). It is built on top of the NASA Land Information System Framework with supports multiple land surface models, data assimilation schemes, and routing schemes (Kumar et al. 2006). Here, the Noah-MP land surface model with multi-parameterization option was used with various physical processes such as runoff generation, dynamic vegetation, canopy stomatal resistance, groundwater, and so on (Niu et al. 2011). For discharge and floodplain inundation, we integrate Noah-MP with the Hydrological Modeling and Analysis Platform (HyMAP) river routing model, which is a state-of-art global scale hydrodynamic model that simulates surface water dynamics in rivers and floodplains using local inertia formulation.

ILDAS was run on a 0.1 degree spatial resolution at a 15 minute timestep with daily outputs from 1981-2021. We use gridded daily rainfall data at 0.25° spatial resolution from the India Meteorological Department (IMD). This gridded data is generated from 6955 gauge stations using an inverse distance weighted interpolation scheme. The other model forcing variables such as radiation and humidity were supplemented from Modern-Era Retrospective Analysis for Research and Applications, Version 2 (MERRA-2; (Gelaro et al. 2017)). The specifications for various ILDAS parameters are given in Table 1.

*Table 1: List of ILDAS components and their specifications*

| ILDAS Component | Specifications |
|---|---|
| Land Surface Model | Noah-MP 4.0.1 |
| Routing Scheme | HyMAP |
| Spatial Extent | 68°-98°E, 5.5°-37.5°N |
| Spatial Resolution | 0.1° |
| Temporal Resolution | 15 minutes Noah-MP 3.6 and HyMAP with adaptive timestep, daily output fields |
| Time Period | 1981-2021 |
| ]Forcing | IMD, MERRA-2, |
| Forcing Variables | Precipitation, near-surface air temperature, near-surface specific humidity, surface pressure, eastward and northward wind velocity, incident longwave and shortwave radiation |
| Forcing Height | 2 m for surface air temperature, specific humidity, and surface pressure, 10 m for wind |
| Topography and river network | MERIT Hydro |
| Soils Definition | (NCAR) STATSGO+FAO blended soil texture map |
| Vegetation Definition | MODIS-IGBP (NCEP-modified), Monfreda et al. (2008) crop types |
| Output Format | NetCDF |

Using a 40-year daily reanalysis data, we estimated the average flooded fraction of a district to use as a robust but static descriptor of flood proneness of the district. To obtain the corrected average flooded fraction and create the flood layer, we subtracted the permanent water layer fraction from the estimated average flooded fraction obtained from ILDAS. The permanent water fraction for each district was calculated from the permanent water layer provided by the Global Surface Water Explorer (GSWE, https://global-surface-water.appspot.com/). The flooded fraction was aggregated to district boundaries, and then the percentage flooded area was calculated with respect to the district area.

### *3.3* Population

The gridded population data at 30-meter resolution was sourced from the Facebook Data for Good platform (https://dataforgood.facebook.com), which provides high resolution population density maps and demographic estimates. This data was re-gridded to the district level to develop latest estimates of population. The impacts of flooding on a district having bigger population will be more harmful compared to another district having smaller population.

### *3.4* Methodology for a District Flood Severity Index (DFSI)

Since a district is the functional administrative unit of India, decisions about disaster management are generally taken at the district level. Decisions about relief and fund disbursement require an accurate country-wide understanding of floods. Thus, we are proposing a District Flood Severity Index (DFSI), with the objective of having a simple index that can be maintained as a spreadsheet by the government. While more sophisticated conceptions of flood severity are possible, the objective here is to develop an index that is easy for the India Meteorological Department (IMD) to maintain in the long term. It is expected that this index would be useful for planning, infrastructure construction, relief and rescue organizations.

We define DFSI as

$$\text{DFSI} = log_{10}\left\{ \begin{matrix} (1 + \mathit{Flood\ events}) \times (1 + \text{Mean Flood Duration}) \times (1 + \mathit{Fatality})^2 \\ \times (1 + \mathit{Injured}) \times (1 + \text{Flooded Area}) \times \text{Population}) \end{matrix} \right\} \quad (1)$$

where,

a) Flood Events = Number of flooding events in a district,
b) Mean Flood Duration = Mean of the flood duration of all flooding events in a district
c) Fatality = Number of human fatalities in a district,
d) Injured = Number of humans injured in a district,
e) Flooded area = Percentage of area of a district that is historically flooded. Calculated as mean of the flooded fraction from the 40 year reanalysis of the Indian Land Data Assimilation System (ILDAS),
f) Population = district population.

Composite DFSI consists of six factors that can be classified in two groups of variables: a) variables representing the occurrence of floods – number of flood events, mean flood duration, flooded area, and population; b) variables representing the damage due to floods – human fatality and the number of humans injured in a district. High weightage is given to fatality and log transformation is used to scale the index in a range which is intuitively easy to visualize.

## 4. Results and Discussions

### *4.1* District and State-wise Floods

The average flooded area of a district as a percentage of total area is given in Figure 2. Some districts in Assam, Bihar, Eastern UP, and isolated districts around the country emerge as districts with highest percentages of their area affected by floods.

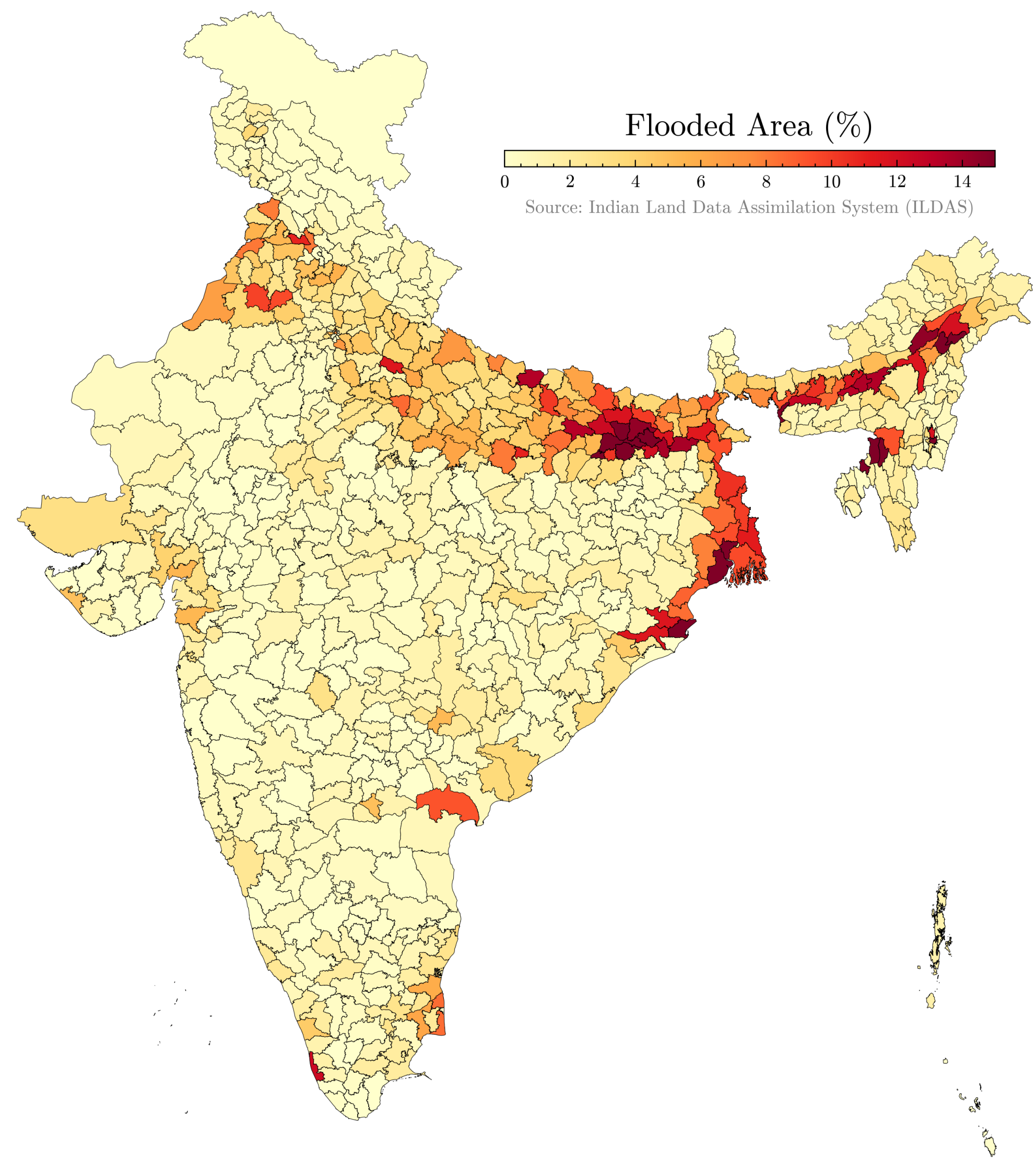

*Figure 2: Average flooded area as percentage of a district's total area calculated from the Indian Land Data Assimilation System (ILDAS) reanalysis*

The total number of flood events per district and per state is presented in Figure 3. The district of Thiruvananthapuram has experienced more than 231 occurrences of floods, or more than 4 events per year on an average. Further, 5 districts had more than 178 occurrences of floods, or more than 3 events per year on an average. These are Thiruvananthapuram, Lakhimpur, Dhemaji, Kamrup, and Nagaon. Also, 28 districts had more than 113 flooding events, or more than 2 events per year. These are Thiruvananthapuram, Lakhimpur, Dhemaji, Kamrup, Nagaon, Mumbai, Barpeta, Sonitpur, Nagpur, Darrang, Dibrugarh, Cachar, Wayanad, Goalpara, Baksa, Pune, Jorhat, Morigaon, Nalbari, Golaghat, Sivasagar, Dhubri, Warangal Rural, Hamirpur,

Idukki, Alappuzha, Kozhikode, and Shimla. Furthermore, 101 districts had more than 56 flood occurrences, which implies that parts of the district faced an average of one event per year. We may categorize districts that experience at least one flood a year as flood prone, and districts that face, on an average, more than three flood events per year to be severely flood prone.

(a)

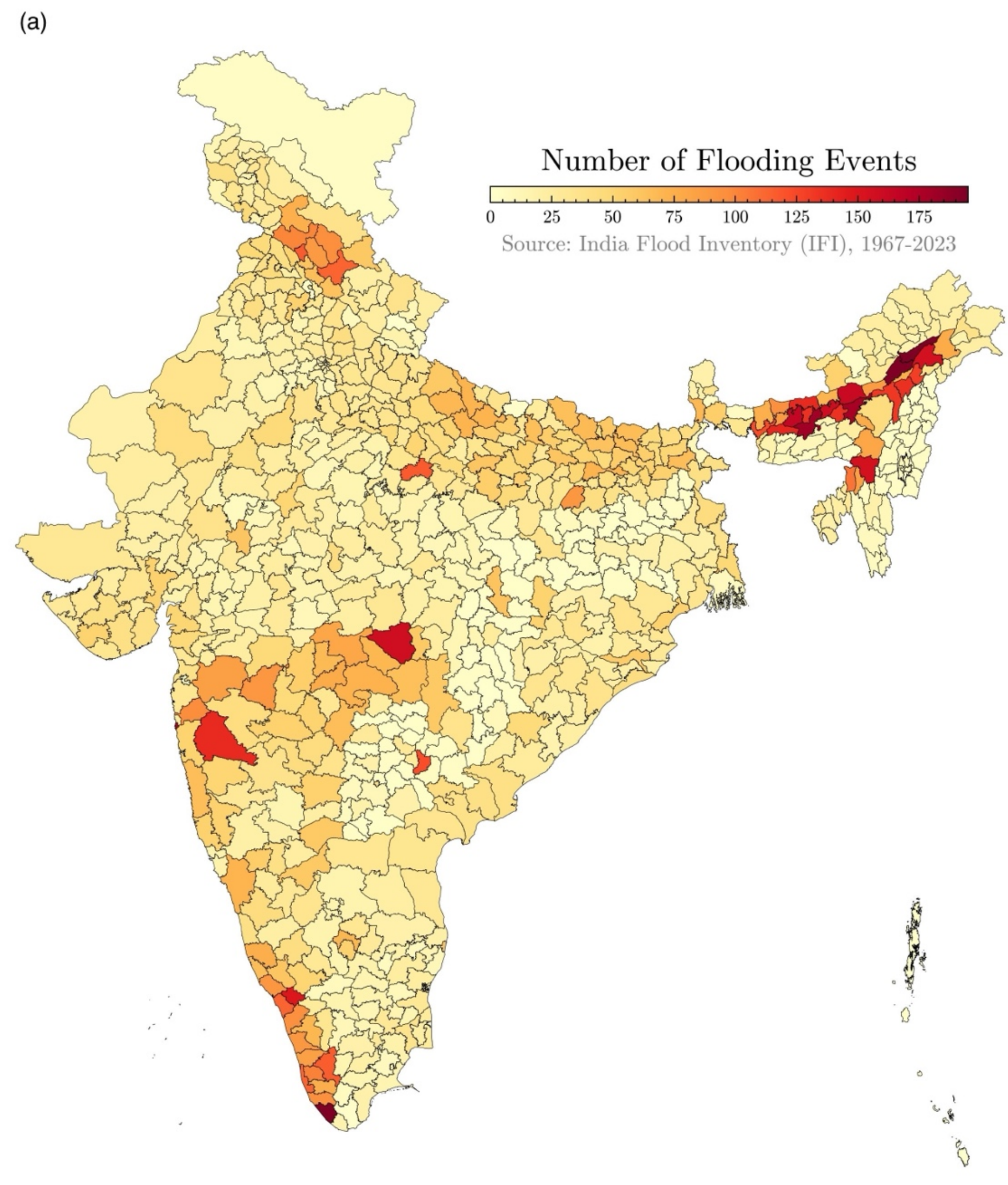

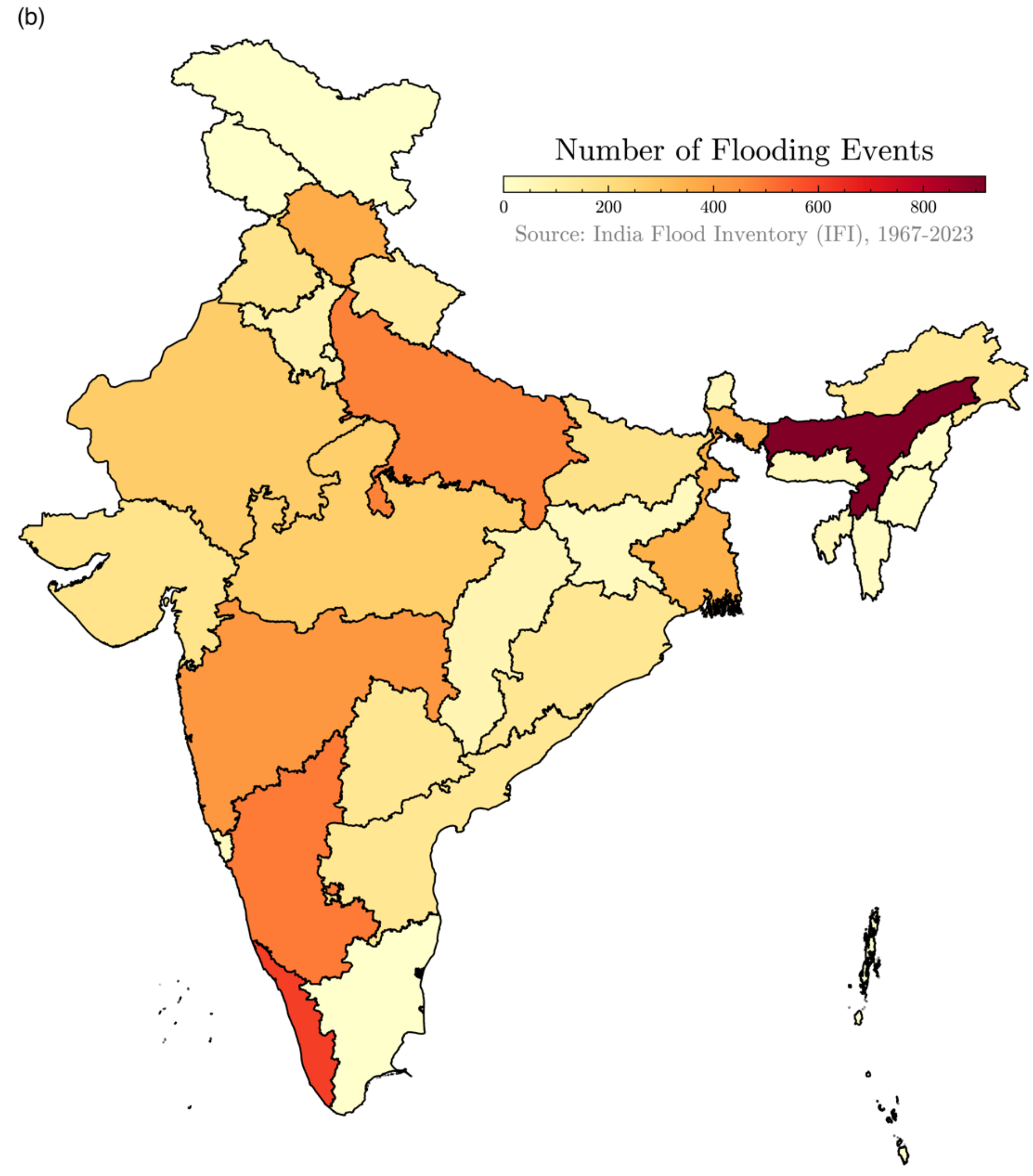

*Figure 3: (a) District wise occurrences of floods in India, and (b) state-wise occurrences of flooding events in India [1967-2023]*

Figure 4(a) shows district wise occurrences of floods in India. It is noted that most flood prone districts are in the Brahmaputra, Ganga basins, and some parts of Maharashtra and Kerala. Figure 4(b) shows state-wise occurrences of flooding events in India [1967-2023] and the States in the North-Eastern region where Brahmaputra and Barak rivers are flowing are severe flood prone.

Expectedly, the state of Assam in North-East India experiences the highest number of floods India, having faced more than 800 flood events over 56 years. This implies that that every year, on an average, the state of Assam is likely to face at least 14 flood events, highlighting the hardships that the people face there, year-after-year. The other flood prone states in descending order of flood events include Kerala, Karnataka, Uttar Pradesh, and Maharashtra.

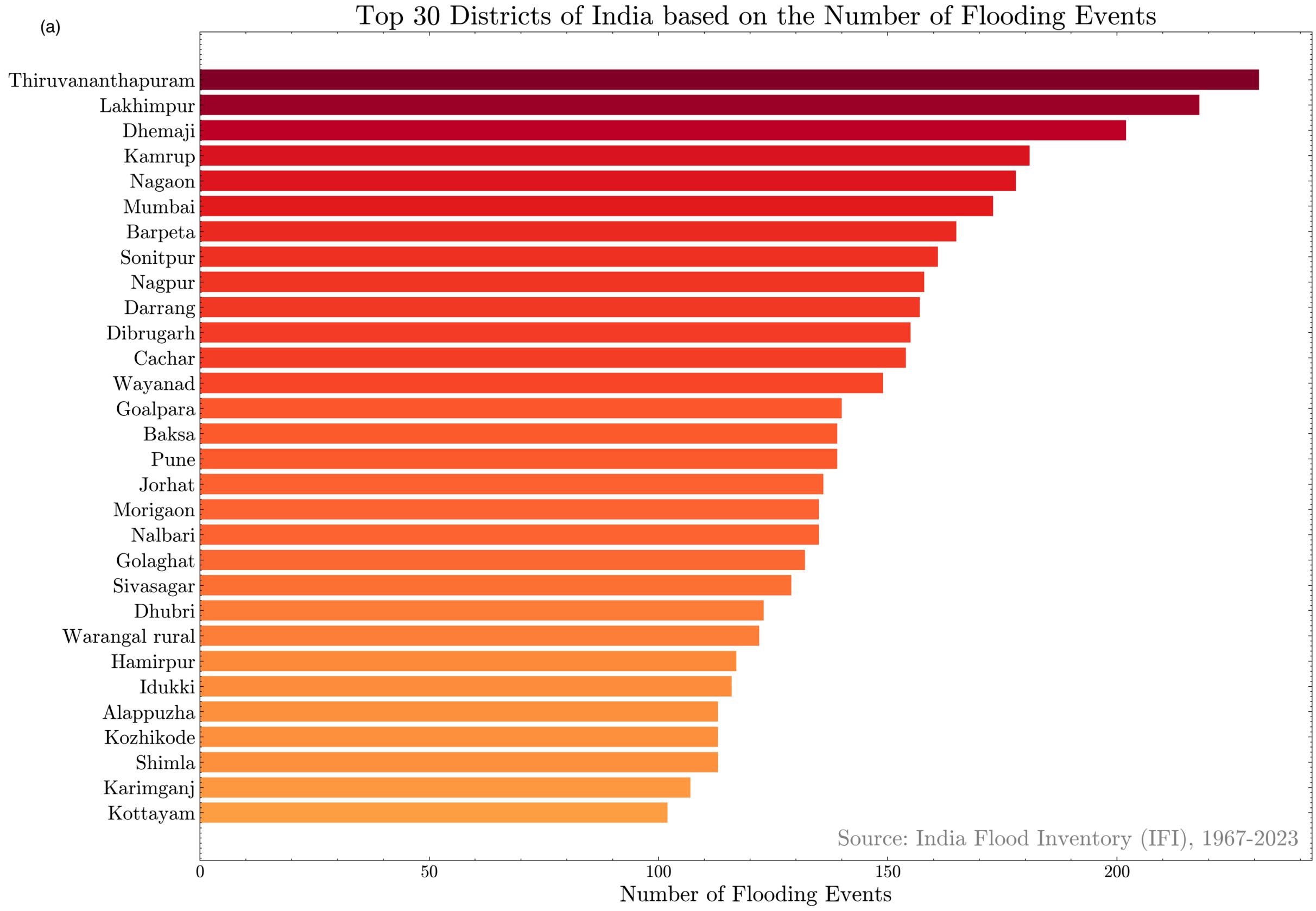

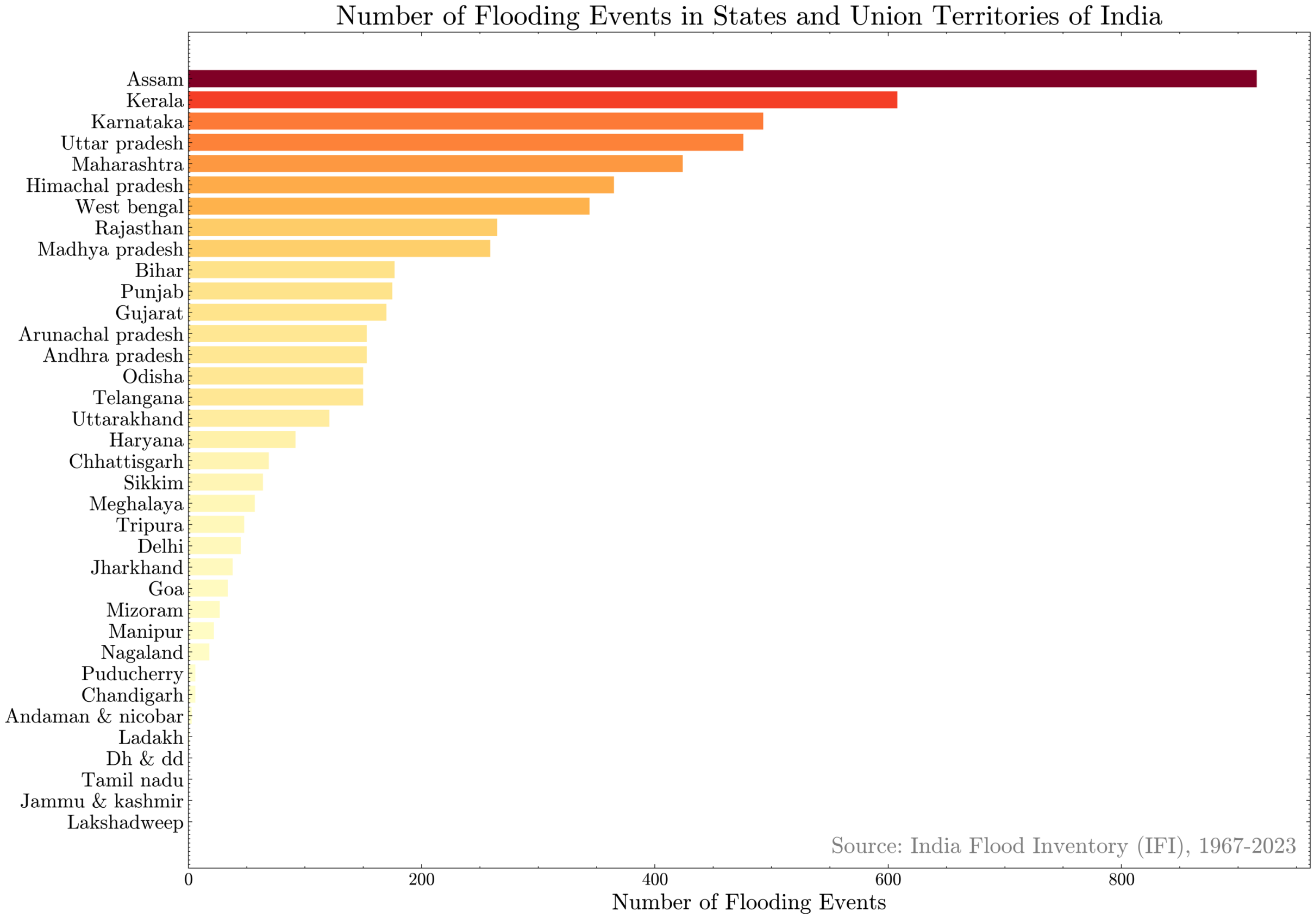

*Figure 4: Top 30 districts in India with the highest number of floods, and (b) State-wise list of highest number of flooding events [1967-2023]*

### *4.2* India Flood Inventory – Death and Damages

The death and damage information in the India Flood Inventory can be considered the best available lower bound of the actual losses as shown in Figure 5. One may note from this figure that due to improved flood management, human fatality is nearly stable or is coming down in recent times despite increase in population in India. The number of deaths is around 1000 per year in recent times and efforts are underway to further reduce this number. Experience from recent flood disasters reveals a few reasons behind large number of deaths are: dwellings unsafe or building homes on unstable hillslopes, houses or too close to rivers or in flood plains, landslides due to intense rainfalls, accentuating flood problem, and crossing rivers in spate without adequate safety gears.

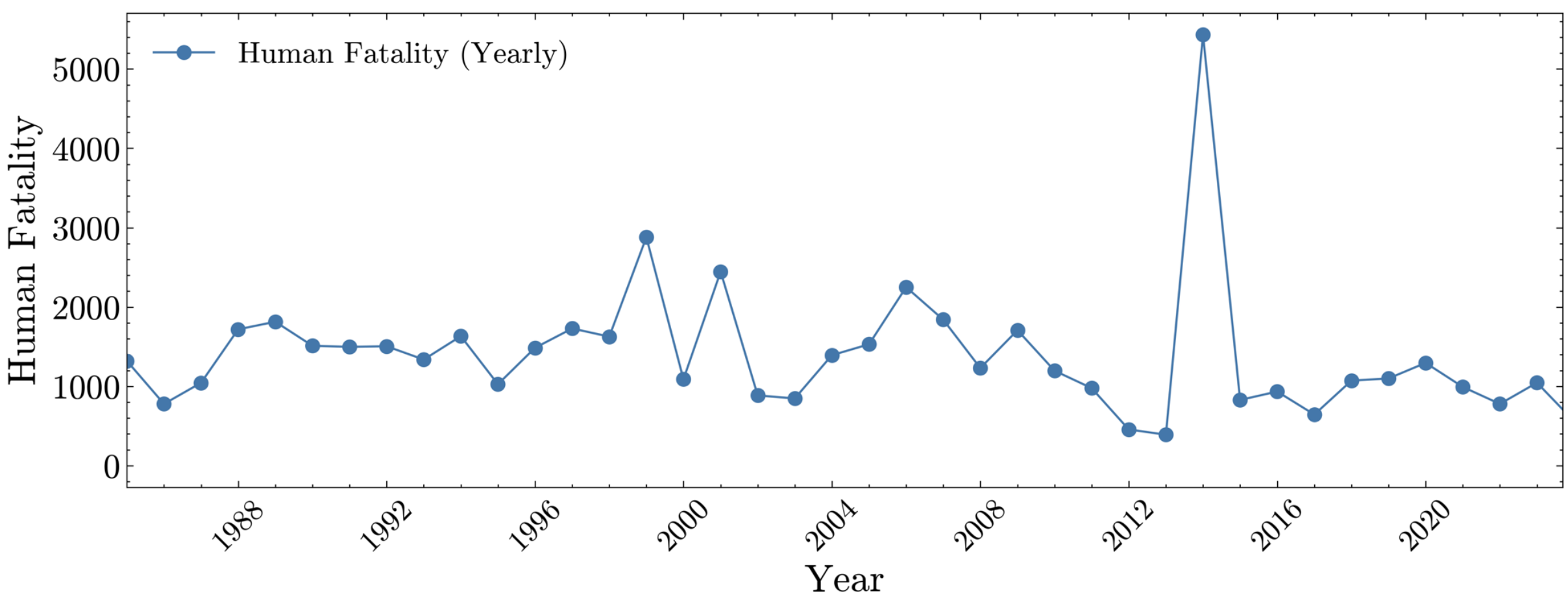

*Figure 5: Evolution of deaths due to flooding events [1967-2023]*

### *4.3* District Flood Severity Index for India

Finally, using Eq. (1), we develop the District Flood Severity Index (DFSI), as given in the Figure 66.

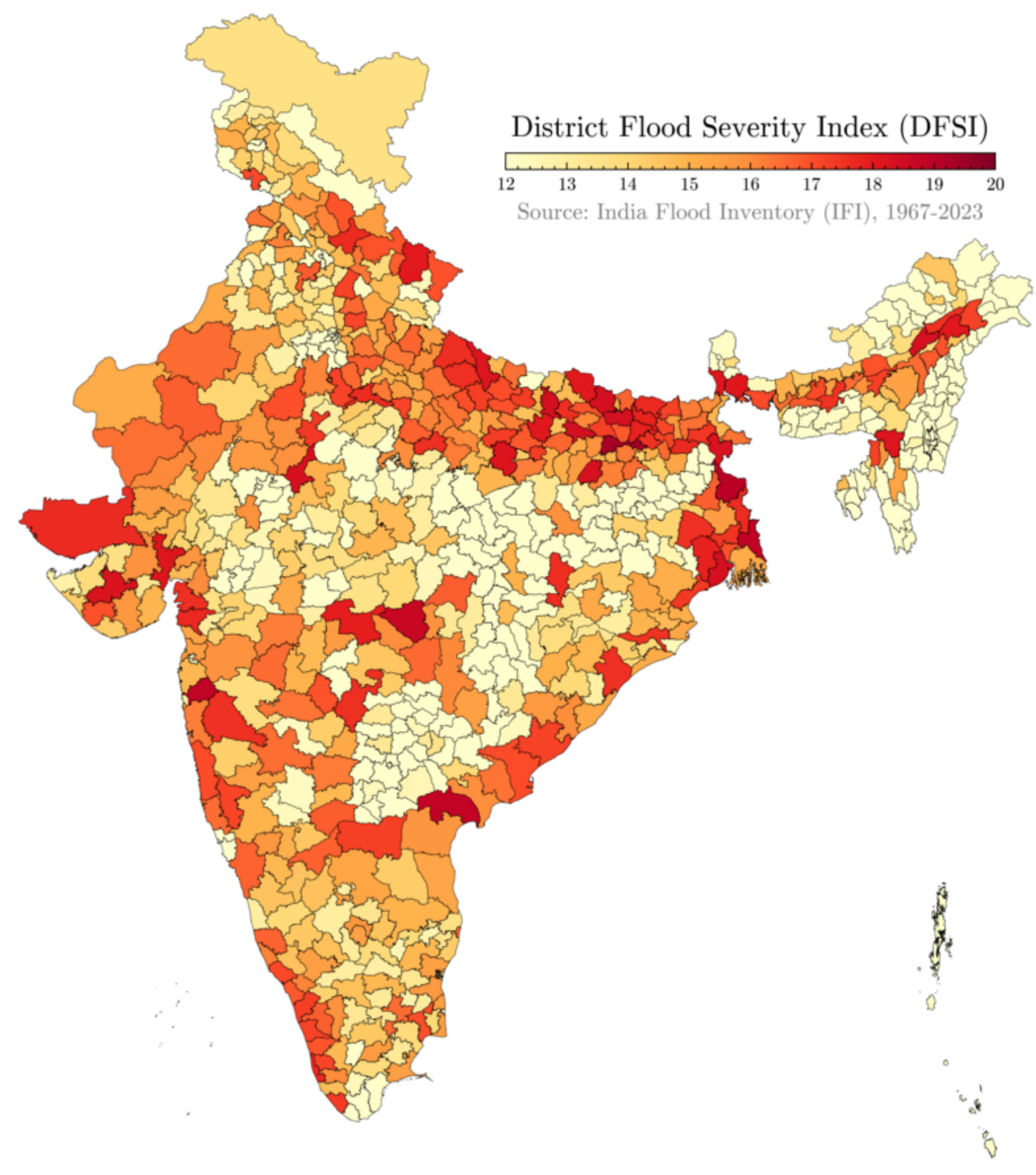

*Figure 6: District Flood Severity Index of India developed using the India Flood Inventory [1967-2023]*

The top 30 districts of India with the highest DFSI are listed in Table 2. Patna (Bihar) emerges as the most affected district in terms of flood severity.

*Table 2: Top 30 districts with the highest District Flood Severity Index*

| DFSI Rank | District | States | DFSI |
|---|---|---|---|
| 1 | Patna | Bihar | 19.37 |
| 2 | Murshidabad | West Bengal | 19.01 |
| 3 | Thane | Maharashtra | 18.88 |

| 4 | North 24 Parganas | West Bengal | 18.86 |
|---|---|---|---|
| 5 | Guntur | Andhra Pradesh | 18.84 |
| 6 | Nagpur | Maharashtra | 18.70 |
| 7 | Gorakhpur | Uttar Pradesh | 18.40 |
| 8 | Ballia | Uttar Pradesh | 18.38 |
| 9 | Purba Champaran | Bihar | 18.38 |
| 10 | Purba Medinipur | West Bengal | 18.37 |
| 11 | Muzaffarpur | Bihar | 18.37 |
| 12 | Lakhimpur | Assam | 18.35 |
| 13 | Kota | Rajasthan | 18.33 |
| 14 | Aurangabad | Bihar | 18.33 |
| 15 | Maldah | West Bengal | 18.28 |
| 16 | Rajkot | Gujarat | 18.25 |
| 17 | Prayagraj | Uttar Pradesh | 18.19 |
| 18 | Aurangabad | Bihar | 18.18 |
| 19 | Bahraich | Uttar Pradesh | 18.17 |
| 20 | Cachar | Assam | 18.16 |
| 21 | Ahmadabad | Gujarat | 18.13 |
| 22 | Jalpaiguri | West Bengal | 18.11 |
| 23 | Darjeeling | West Bengal | 18.09 |
| 24 | Dibrugarh | Assam | 18.08 |
| 25 | Azamgarh | Uttar Pradesh | 18.07 |
| 26 | Chamoli | Uttarakhand | 18.06 |
| 27 | Pashchim Champaran | Bihar | 18.02 |
| 28 | Amravati | Maharashtra | 17.92 |
| 29 | Medinipur West | West Bengal | 17.87 |
| 30 | Samastipur | Bihar | 17.85 |

## 5. Discussion

A perusal of Table 2 reveals that most districts with high DFSI are either by side of a river or on coast and this fact provides guidance for flood management in India. In those cases where flooding occurs due to high river flows, the problem can be partly addressed by conserving and utilizing flood flows (Jain 2021). Attempts should be made to conserve as much flood flows as feasible in upstream reservoirs/ponds and the river improvements need to be performed to safely pass the remaining flows. In case of coastal floods, the best options are early warning systems, climate resilient developments, and preparedness.

The Chamoli district is a hilly district in Uttarakhand. It does not face recurrent flood problem but appears in this list because of a few isolated highly damaging flood events. Further, among the list of 30 districts, 17 are in the Ganga basin and 3 in the Brahmaputra basin. Among all the Indian river basins, human population is highest in the Ganga basin and high flood proneness of this basin is worrisome for professionals and decision makers. Additionally, Brahmaputra is a transboundary basin that affects four countries, and comprehensive flood mitigation in Brahmaputra river basin would require national efforts and regional cooperation.

Because of high population pressure, some people live too close to the river which also provides them their livelihood. But these people are the first sufferers when the floods strike. To reduce deaths and injuries due to floods, it would be necessary to move these people to safer locations. The concept "room for the river" has been successfully practiced in the Netherlands and may be implemented in India as well. The concept "room for the river" towards integrated flood risk management is successfully implemented "on the ground" in Netherlands (Verweij et al. 2021). The concept embraces a multi-functional river utilizing a holistic approach in which flood safety is realized in combination with other values such as landscape, environmental and cultural values (Zevenbergen et al. 2015). The instrument may prove valuable for integrated flood risk management practices in India.

Thereby based on the DFSI maps, critical infrastructure in the districts may be identified by infrastructure managers which may be affected by floods. Integrating the data with flood hazard, vulnerabilities and exposure data, a comprehensive quantitative risk analysis may be carried out by authorities towards business continuity planning (BCP). The BCP approach provides for safety and reliability of infrastructure services by avoiding disaster losses with focus on managing risks rather than managing disasters by assessment of repair costs of critical infrastructure vis-à-vis damage costs of critical infrastructure during disasters (Joshi et al. 2024).

## 6. Conclusions and Future Work

Analysis of data compiled in this study has confirmed that floods are the worst natural disaster that continues to hit India repeatedly. The flood problem is likely to become more severe due to climate and land use/cover changes. The present work has created, for the first time in India, a DFSI at a fine spatial scale by using data pertaining to six indicators representing flood occurrences and damage caused. We have computed DFSI at the district level which is an administrative unit for planning and development activities. Further, the data used here was available at this scale and we have employed GIS and other modern tools to process and store the data. We hope that this index would be useful for planning and decision making. Based on time series of DFSI maps prepared for different years would reveal trends in flooding and better targeted flood management.

To take this work forward, better data on more variables would be necessary. The authors did not have access to data about economic damages due to floods. Such data are collected for providing compensation to the victims of floods but comprehensive economic loss data at finer resolutions is not easily accessible. The spatial data about floods needs to be collected by laying down protocols. Use of remote sensing data along with ground truthing (to verify values and to fill gaps) and storage of this data in modern geospatial databases would make it easier to analyze the data. We hope that this index would be useful for planning and decision making. Based on time series of DFSI maps prepared for different years would reveal trends in flooding and better targeted flood management.

**Acknowledgements**

This research was conducted in the HydroSense lab (https://hydrosense.iitd.ac.in/) of IIT Delhi and the authors acknowledge the IIT Delhi High Performance Computing facility for providing computational and storage resources. Dr. Manabendra Saharia gratefully acknowledges financial support for this work through grants from Ministry of Earth Sciences/IITM Pune Monsoon Mission III (RP04574) and Ministry of Earth Sciences DeepINDRA Project (RP04741). The authors gratefully acknowledge the Central Water Commission (CWC), National Water Informatics Centre (NWIC), and the Ministry of Jal Shakti (MoJS) for providing the streamflow datasets used in this study. Authors gratefully acknowledge the Indian Meteorological Department (IMD) for providing access to the datasets.

**Compliance with Ethical Standards**

The authors declare that they have no known competing financial interests or personal relationships that could have appeared to influence the work reported in this paper.

**Author Contributions**

MS: Conceptualization, Writing, Editing

SJ: Conceptualization, Writing, Editing

VP: Analysis

HM: Data Curation

OPS: Original datasets

DJ: Editing

**Data Availability**

- The India Flood Inventory with Impacts (IFI-Impacts) database is published in Zenodo: https://doi.org/10.5281/zenodo.11069835
- Analysis notebooks: https://github.com/msaharia/India-flood-inventory-impacts